\documentclass[11pt]{article}

\usepackage[mathscr]{euscript}

\usepackage{amsmath,amsfonts,amsbsy,amssymb,array,accents,dsfont}
\usepackage{enumerate,array,latexsym,graphicx,mathrsfs,verbatim,psfrag}
\usepackage{bm} 
\usepackage[normalem]{ulem}
\usepackage{multirow}
\usepackage{cellspace}
\usepackage{pdflscape}
\usepackage{booktabs} 
\usepackage[usenames]{color}
\usepackage[utf8]{inputenc}
\usepackage[english]{babel}
\usepackage{fancybox}

\usepackage{datetime}

\usepackage[nosort]{cite}
\usepackage{chngpage} 
\usepackage{setspace}
\usepackage{tensor}
\usepackage{bbold}

\usepackage{enumitem}
\setlist{itemsep=0pt}

\usepackage{geometry}
\usepackage{longtable} 
\allowdisplaybreaks
\usepackage{pdflscape}

\usepackage[vcentermath,stdtext]{youngtab} 
\usepackage{lieart} 

\definecolor{cardinal}{rgb}{0.6,0,0}
\definecolor{darkgreen}{rgb}{0.1,0.4,0}
\definecolor{golden}{rgb}{0.92, 0.7, 0}
\definecolor{midnight}{rgb}{0, 0, 0.5}
\definecolor{darkblue}{rgb}{0.3,0.3,0.7}
\definecolor{darkred}{rgb}{0.7,0,0}

\usepackage{dsfont}

\usepackage{hyperref}
\hypersetup{
  colorlinks=true,
  linkcolor=darkblue,
  citecolor=darkred,
  urlcolor=darkgreen,
  linktocpage=true,
  linktoc=page
}

\usepackage{arydshln}
\usepackage{mathtools}

\newcommand{\captionfonts}{\small}
\makeatletter  
\long\def\@makecaption#1#2{%
  \vskip\abovecaptionskip
  \sbox\@tempboxa{{\captionfonts #1: #2}}%
 \ifdim \wd\@tempboxa >\hsize
    {\captionfonts #1: #2\par}
  \else
    \hbox to\hsize{\hfil\box\@tempboxa\hfil}%
  \fi
  \vskip\belowcaptionskip}
\makeatother   

\DeclareMathSymbol{\medhatsym}{\mathord}{largesymbols}{"62} 

\DeclareMathSymbol{\medtildesym}{\mathord}{largesymbols}{"65}

\newcommand{\comm}[1]{} 

\def\({\left(}
\def\){\right)}
\def\[{\left[}
\def\]{\right]}

\def\One{{\hbox{ 1\kern-.8mm l}}}

\def\barray{\begin{array}}
\def\earray{\end{array}}
\def\be{\begin{equation}}
\def\ee{\end{equation}}
\def\bea{\begin{eqnarray}}
\def\eea{\end{eqnarray}}
\def\bal{\begin{align}}
\def\eal{\end{align}}

\numberwithin{equation}{section} 
\numberwithin{table}{section} 

\newenvironment{al}
    {\begin{equation}\begin{aligned}}
    {\end{aligned}\end{equation}}
\newcommand{\f}[2]{\frac{#1}{#2}}

\makeatletter
\g@addto@macro\bfseries{\boldmath}
\makeatother

\definecolor{cardinal}{rgb}{0.6,0,0}
\definecolor{darkgreen}{rgb}{0,0.4,0}
\definecolor{purple}{rgb}{0.5, 0, 0.5}
\definecolor{golden}{rgb}{0.92, 0.7, 0}
\definecolor{midnight}{rgb}{0, 0, 0.5}
\definecolor{darkblue}{rgb}{0, 0, 0.8}

\def\cB{{\cal B}}

\def\cD{{\cal D}}

\newcommand{\abs}[1]{\ensuremath{\left|#1\right|}}

\begin{document}


\begin{flushright}
\end{flushright}

\vspace{6mm}

\begin{center}
\begin{adjustwidth}{-7mm}{-7mm} 
\begin{center}
{\fontsize{20}{20} \bf{Asymptotically flat $(1,m,n)$ superstrata: \\ a farewell to AdS   }} \medskip \\
\end{center}
\end{adjustwidth}
\vspace{10mm}

\centerline{{\bf Willem Govaerts, Robert Walker}}
\bigskip
\bigskip
\vspace{1mm}

\centerline{Instituut voor Theoretische Fysica, KU Leuven,} 
\centerline{Celestijnenlaan 200D, B-3001 Leuven, Belgium}

\vspace{9mm} 
{\footnotesize\upshape\ttfamily willem.govaerts@outlook.be ,~ robert.walker @ kuleuven.be} \\

\vspace{9mm}
 
\textsc{Abstract}

\end{center}
\begin{adjustwidth}{6mm}{6mm} 
 
\vspace{2mm}
\noindent
\noindent
Superstrata microstate geometries furnish some of the most successful laboratories, to date, for probing black hole microstructure in a geometric setting. This paper extends the $(1,m,n)$ family of superstrata, to allow for flat R$^{1,5}$ asymptotics. Previous constructions utilized the decoupling regime, where the geometry is asymptotically AdS$_3 \times $S$^3$. Brief comments are made on the additional complexity introduced by the flat space coupling, how this obscures holomorphicity and breaks the consistent truncation to which the decoupled solutions belong. Holomorphicity and consistent truncation were key simplifications assisting previous studies of the $(1,m,n)$ superstrata, undertaken in the decoupling regime. Further, these results open a window for future projects to determine how previous analysis on the decoupled geometries extend or are modified once flat space asymptotics are imposed. Since our universe is almost flat on cosmological scales, this represents progress towards more phenomenologically relevant microstate geometries. This work can be considered a continuation of that in \cite{Bena:2017xbt}, where some single mode superstrata were also coupled to flat space.    

\bigskip

\end{adjustwidth}

\vspace{8mm}
 

\thispagestyle{empty}

\newpage


\baselineskip=11.5pt
\parskip=1pt
\setcounter{tocdepth}{2}

\tableofcontents

\newpage 

\baselineskip=15pt
\parskip=3pt

\section{Introduction}
\label{sect:introduction}

In the search for a satisfactory description of black hole microstates, \textit{microstate geometries} furnish some of the most readily accessible and instructive laboratories. These geometries provide families of smooth supergravity solutions, that can arbitrarily closely approximate the geometry of a black hole outside the horizon, before capping off smoothly at the would be horizon scale in a multitude of ways. Historically, the D1/D5/P system of IIB supergravity has been particularly fruitful for constructing new microstate geometries \cite{Bena:2006kb,Bena:2007qc,Bena:2008wt,Bena:2010gg,Bena:2015bea,Bena:2016ypk,Heidmann:2017cxt,Bena:2017geu,Avila:2017pwi,Bena:2017xbt,Heidmann:2019xrd,Ganchev:2021pgs,Ganchev:2021iwy, Ceplak:2022pep}, see \cite{Bena:2007kg,Warner:2019jll,Bena:2022ldq} for convenient reviews. Additionally, working in the \textit{decoupling} regime, where the geometry is asymptotically AdS, allows insights from field theory to be leveraged, via application of the AdS/CFT correspondence, simplifying many analysis.    

Although the decoupling limit is extremely useful for building and probing microstate geometries, it is at odds with the asymptotic geometry of black holes in our universe. Experimentally, it is well established that on cosmological scales our universe is very close to being flat \cite{Spergel_2003,Spergel_2007}. Thus, it is important to establish:

\textit{ What features of microstate geometries, in the decoupling limit, are altered or retained when more realistic asymptotics are imposed?} 

This paper addresses this question, by explicitly coupling a previously known (in the decoupling limit) family of microstate geometries to flat space. Then analyzing the resulting novel features, that may be conceptualised as ``corrections" to the decoupled solutions. It should be noted that although this represents progress toward more realistic asymptotics, the superstrata considered here are six-dimensional. Thus, our results are not directly phenomenologically applicable, but may provide indicative qualitative results of relevance to phenomenology. 

The exemplar for the exploitation of AdS/CFT, for the construction of microstate geometries, are the \textit{superstrata} solutions \cite{Bena:2015bea,Bena:2016agb,Bena:2016ypk,Bena:2017xbt,Bakhshaei:2018vux,Ceplak:2018pws,Heidmann:2019zws,Shigemori:2019orj,Heidmann:2019xrd,Mayerson:2020tcl,Shigemori:2020yuo,Mayerson:2020acj,Ganchev:2021iwy}. These solutions are dual to specific states in the D1/D5 CFT, labelled by three integer mode numbers $(k,m,n)$, coming in two types\footnote{There are now more recent variants, that were not fully considered in the context of this paper, such as the vector-superstrata of \cite{Ceplak:2022pep}.}: the original, \cite{Bena:2015bea, Bena:2016ypk, Bena:2017xbt}, and the supercharged, \cite{Ceplak:2018pws}. Further, these states can be superposed, so that a generic superstratum solution is specified by the choice of two functions of three variables, accounting for superposition of all possible modes of each type. The appropriate superpositions are constrained by demanding regularity of the final geometry, in a procedure which has come to be known as \textit{coiffuring}. Further, the superstrata have been exploited in many precision holographic calculations \cite{Giusto:2015dfa,Galliani:2016cai,Bombini:2017sge,Bombini:2019vnc,Tian:2019ash,GarciaTormo:2019inl,Giusto:2019qig,Bena:2019azk,Giusto:2019pxc,Hulik:2019pwr,Giusto:2020mup,Ceplak:2021wzz,Rawash:2021pik}, as well as purely within supergravity, to shed light on potential resolutions to the information paradox \cite{Hawking:1975vcx,Mathur:2009hf}. 

Unfortunately, due to technical limitations, the full analytic expression for a generic superstratum solution of two functions of three variables, is not known. Instead, subfamilies are known and are indexed by the $(k,m,n)$ modes being considered in the superposition, see \cite{Heidmann:2019xrd} for examples. An instructive family is the $(1,m,n)$ solutions, constructed in \cite{Mayerson:2020tcl}. This family rose to prominence due to the separability of the massless wave equations of its $(1,0,n)$ and $(1,1,n)$ subfamilies \cite{Bena:2017upb,Walker:2019ntz}, which allowed many explicit calculations to be performed. This separability was ultimately traced to the remarkable fact that the $(1,m,n)$ family belongs to a consistent truncation on a deformed three-sphere \cite{Pilch:2021ndc}, and the deformations of the three sphere preserve some ``nice isometries".   

A road map for coupling superstrata to flat space was developed in \cite{Bena:2017xbt}, then applied to some single mode original superstrata. In this paper we apply this procedure to find the full $(1,m,n)$ asymptotically flat multimode superstrata. The resulting geometries are vastly more involved, can no longer be written cleanly in the holomorphic variables identified in \cite{Heidmann:2019xrd} and fall outside the consistent truncation developed in \cite{Mayerson:2020tcl}. Thus, many of ``nice features" that aided in the construction and probing of the decoupled $(1,m,n)$ superstrata, disappear once they are coupled back to flat space. This raises the question:

\textit{To what extent are the currently accepted results or intuition, in the microstate geometry program, artefacts of working in the decoupling limit?}

Beyond identifying the aforementioned disappearance of nice features and identifying the potential for analysing UV/IR mixing within the microstate geometry program, a more complete analysis of this question is left to a future study. It is entirely  possible that most of the intuition developed in the decoupling limit carries over, just the technical details become an order of magnitude more complicated.  

The remainder of this paper is organized as follows:
\begin{itemize}
\item Section \ref{sect:6Dn1} provides an introduction to the six-dimensional supergravity, in which the superstrata are most easily studied. Further, the foundational semi-linear form of the BPS equations, which aid solution construction, are presented. 
\item Section \ref{sect:full1mn} presents the full $(1,m,n)$ solutions and demonstrates their regularity at key locations. There are sufficient parameters in this solution to reproduce the decoupled family as well as the asymptotically flat family. In addition, the technical details involved in this construction are collected in Appendix \ref{app:Constructing1mn}. 
\item Section \ref{sect:1mnAdSSolutions} presents the decoupled $(1,m,n)$ solutions, that are asymptotically AdS$_3 \times$S$^3$. Introducing the details of both the holomorphic variables and consistent truncation.
\item Section \ref{sect:TheGeometry} illustrates the geometry of the asymptotically flat $(1,m,n)$ geometries in three illuminating limits: the flat asymptotic region, the smooth cap at the supertube locus and the long BTZ like throat regime.
\item Finally, in section \ref{sect:Conclusions} we discuss the results of this paper and make some concluding remarks on future directions. 
\end{itemize}

\section{Six-dimensional $\mathcal{N}=1$ supergravity with two tensor multiplets}
\label{sect:6Dn1}

This section presents the six-dimensional supergravity theory to which the superstrata belong. Further, the BPS equations are given in their remarkable semi-linear form. It is this semi-linear form that is at the heart of the solution generating procedures, which allows for such involved microstate geometries to be constructed analytically.

\subsection{The theory}
\label{ss:thetheory}

The superstrata are most simply constructed and studied in six-dimensional, $\mathcal{N}=(1,0)$ supergravity, coupled to two tensor multiplets, identified in \cite{Bena:2015bea,deLange:2015gca,Mayerson:2020tcl}. The bosonic field content of this theory can be organized into: the metric $g_{\mu\nu}$, two scalars $(\varphi,X)$, and three three-forms which we label by $G^{I}$ with $I\in \left\lbrace 1,2,4 \right\rbrace$. It is convenient to introduce the $SO(1,2)$ metric, $\eta^{IJ}$, with non-zero components:
\begin{align}
\eta^{12}=\eta^{21}=1 \qquad \text{and} \qquad \eta^{44}=-2\,, \label{etaDef}
\end{align}
which is used to raise and lower $I,J,\cdots$ indices.

The equations of motion for the three forms are given by the twisted self duality relations:
\begin{align}
*G^{I} = \tensor{M}{^{I}_{J}} G^{J}\,, \label{TwistedSD}
\end{align}
where
\begin{align}
\tensor{M}{^{I}_{J}} = \frac{1}{2}e^{\sqrt{2}\varphi} \begin{pmatrix}
X^{2} & 8 & -2\sqrt{2}X \\
\frac{e^{-2\sqrt{2}\varphi}}{8}\left( 2+e^{\sqrt{2}\varphi}X^{2} \right)^{2} & X^{2} & -\frac{X}{2\sqrt{2}}\left( 2e^{-\sqrt{2}\varphi}+X^{2} \right) \\
 \frac{X}{\sqrt{2}}\left(2e^{-\sqrt{2}\varphi} +X^{2} \right) & 4\sqrt{2}X & -2e^{-\sqrt{2}\varphi} - 2X^{2}
\end{pmatrix}\,,
\end{align}
together with the Bianchi identities
\begin{align}
dG^{I} = 0\,.
\end{align}

The remaining equations of motion can be obtained by minimizing the action constructed from the Lagrangian
\begin{align}
\mathcal{L} = \left[ R - \frac{1}{2}g^{\mu\nu}\left( \partial_{\mu}\varphi \partial_{\nu}\varphi +e^{\sqrt{2}\varphi}\partial_{\mu}X\partial_{\nu}X \right)- \frac{1}{6}M_{IJ}\tensor{G}{^{I}_{\mu\nu\rho}}G^{J\mu\nu\rho}\right] *1 \,.
\end{align}

\subsection{Useful parametrization of bosonic field content}
\label{ss:paramboson}

It turns out that the BPS equations, consistent with the supersymmetries of the D1/D5/P system, take a semi-linear form when written in terms of a very specific parametrization of the bosonic fields introduced in the previous subsection, \cite{Bena:2011dd}. One introduces sets of: functions $(Z_{I},\mathcal{F})$, one forms $(\beta,\omega)$ and two forms $\Theta^{I}$. Upon the specification of a hyper-K\"{a}hler base $ds_4^2(\mathcal{B})$, the BPS equations reduce to solving for this \textit{BPS data}: $(\beta,Z_{I},\Theta^{I},\mathcal{F},\omega)$.

The most general six-dimensional metric, consistent with the (super)symmetries of the D1/D5/P system, was considered in \cite{Gutowski:2003rg}:
\begin{align} \label{ds6Intro}
ds_6^2 &=   -\frac{2}{\sqrt{\mathcal{P}}} \, (dv+\beta) \big(du +  \omega + \tfrac{1}{2}\, \mathcal{F} \, (dv+\beta)\big) 
+  \sqrt{\mathcal{P}} \, ds_4^2(\mathcal{B})\,. 
\end{align} 
In this expression, $(u,v)$ are light cone coordinates related to the $y$-circle\footnote{This is the common compact direction that the D1 and D5 branes wrap.} (of circumference $2\pi R_{y}$) by:
\begin{align} 
u =\frac{1}{\sqrt{2}}(t-y) \qquad \text{and} \qquad v= \frac{1}{\sqrt{2}}(t+y)\,. \label{uvDefUlt}
\end{align} 
The function $\mathcal{P}$ is given in terms if the $Z_{I}$ via:
\begin{align}
\mathcal{P} = \frac{1}{2}\eta^{IJ}Z_{I}Z_{J}\,.
\end{align}
While, $(\beta,\omega)$ are one forms on $\cB$, and all the BPS data may have functional dependence on every coordinate except $u$.

The three forms $G^{I}$, can be expanded in terms of a set of ``electric" functions $Z^{I}$ and ``magnetic" two forms $\Theta^{I}$, as:
\begin{align}
G^{I} =  d \left[ -\frac{1}{2}\frac{\eta^{IJ}Z_{J}}{\mathcal{P}}(du+\omega)\wedge (dv+\beta) \right] + \frac{1}{2}\eta^{IJ}*_{4}\mathcal{D}Z_{J}+\frac{1}{2}(dv+\beta)\wedge \Theta^{I}\,. \label{GdefZTheta}
\end{align}
Note that we use $(d,*)$ to refer to the exterior derivative and Hodge star with respect to the full six-dimensional geometry, (\ref{ds6Intro}), while $*_{4}$ (and in the following $d_{4}$) are with respect to $ds_4^2(\mathcal{B})$. 

The remaining bosonic fields $(\varphi, X)$ can also be expressed in terms of this BPS data, see \cite{Giusto:2013rxa,deLange:2015gca} for further details.

\subsection{BPS equations}
\label{ss:BPSequations}

The BPS-equations can be arranged and solved in the \textit{semi-linear}, layered form:
\begin{itemize}
\item Fix a hyper-K\"{a}hler base $ds_4^2(\mathcal{B})$, and choose a $\beta$ satisfying: 
\begin{align}
d_{4}\beta = *_{4}d_{4}\beta \,.\label{BPS6DLayer0}
\end{align}
\item Find $(Z_{I},\Theta^{I})$ that solve the \textit{first layer}:
\begin{align}
*_{4}\mathcal{D}\dot{Z}_{I} = -\eta_{IJ}\mathcal{D}\Theta^{J} \,, \qquad \mathcal{D}*_{4}\mathcal{D}Z_{I}=\eta_{IJ} \Theta^{J}\wedge d\beta \,, \qquad \Theta^{I} = *_{4} \Theta^{I}\,. \label{BPS6DLayer1}
\end{align} 
\item Find $(\mathcal{F},\omega)$ that solve the \textit{second layer}:
\begin{align}
(1+*_{4})\mathcal{D}\omega +\mathcal{F}\, d_{4}\beta &= Z_{I}\Theta^{I} \,, \label{BPS6DLayer2a}\\
*_{4}\mathcal{D}*_{4} \left(\dot{\omega}- \frac{1}{2}\mathcal{D}\mathcal{F} \right) &= -\frac{1}{4}\eta_{IJ}\left[ 4\ddot{Z}^{I}Z^{J}+2\dot{Z}^{I}\dot{Z}^{J} - *_{4}\left( \Theta^{I}\wedge \Theta^{J} \right)\right]\,. \label{BPS6DLayer2b}
\end{align}
\end{itemize}

This system is semi-linear, in the sense that, once the base geometry is fixed, if one finds a set of \textit{basic data}, $(z_{k,m,n},\vartheta_{k,m,n})$, satisfying:
\begin{equation}
*_4\cD\,\dot{{z}}_{k,m,n} = -\cD\,{\vartheta}_{k,m,n}, \qquad \cD *_4 \cD\, {z}_{k,m,n} =  {\vartheta}_{k,m,n} \wedge d\beta,\qquad {\vartheta}_{k,m,n} = *_4 {\vartheta}_{k,m,n}~, \label{SuperpositionModes}
\end{equation}
then their superposition:
\begin{align}
Z_{I}=\sum_{k,m,n}b_{k,m,n}z_{k,m,n} \qquad \text{and} \qquad \Theta_{I}=\sum_{k,m,n}b_{k,m,n}\vartheta_{k,m,n}~,
\end{align}
with constants, $b_{k,m,n}$, will also solve the first layer (\ref{BPS6DLayer1}). Once the first layer solution is fixed, the second layer differential equations, (\ref{BPS6DLayer2a})-(\ref{BPS6DLayer2b}), are also linear in $(\mathcal{F},\omega)$. The simplification afforded by organizing the BPS equations in this manner is significant. A priori, the non-linear interactions between the base, $\mathcal{B}$, and the data $(\beta, Z_{I},\Theta^{I},\mathcal{F},\omega)$, is extremely difficult to handle when attempting to find solutions.

\subsection{Gauge freedom}
\label{ss:GaugeFreedom}

There is an important gauge transformation of the BPS equations, given by the residual ability to redefine the $u$ coordinate:
\begin{equation} 
u\to  u + U(x^i,v) \qquad \Leftrightarrow \qquad  \omega \to  \omega - d_4U + \dot{U} \,\beta \,,\quad \mathcal{F} \to \mathcal{F} - 2 \dot{U}\,,
\label{gaugetrf}
\end{equation}
where the $x^i$ are coordinates on $\mathcal{B}$. It is straightforward to check this transformation leaves the six-dimensional metric (\ref{ds6Intro}) invariant, while solutions of the BPS equations, (\ref{BPS6DLayer0})-(\ref{BPS6DLayer2b}), are mapped again to solutions under (\ref{gaugetrf}). In order to illuminate certain aspects of microstate geometries, such as their asymptotics, or aid in solution generation, one may need to identify a convenient gauge.

\section{The full $(1,m,n)$ superstrata solutions}
\label{sect:full1mn}

This section presents the full $(1,m,n)$ superstrata solutions, representing the main technical result of this paper. These solutions possess enough tunable parameters to produce both the decoupled asymptotically AdS$_3 \times$S$^3$ family, as well as the asymptotically, R$^{1,5}$, flat family. The technical details of how these solutions were constructed, by weaving together partial results in the literature and novel solutions of the BPS equations, can be found in Appendix \ref{app:Constructing1mn}.

\subsection{``Basic data"}
\label{ss:BasicData}

The superstrata, in six-dimensions, utilize the simplest hyper-K\"{a}hler base, namely flat space. However, it is convenient to adapt the coordinates to the ring of the underlying \textit{supertube}\footnote{The supertube solution is given by setting all $b_{k,m,n}=0$ in the following presentation. The superstrata were originally conjectured to exist in \cite{Bena:2011uw}, via so called ``supertube transitions," on top of this basic supertube solution. These transitions amounts to allowing density variations of the underlying branes and momentum in the D1/D5 frame.}. This is done by introducing \textit{spherical bipolar} coordinates, $(r,\theta,\varphi_{1},\varphi_{2})$, in terms of which\footnote{Note that these spherical bipolar coordinates are related to a standard Cartesian coordinate system $(y^{1},y^{2},y^{3},y^{4})$, for R$^4$, via:
\begin{equation*}
y^{1} + i y^{2} = \sqrt{a^{2}+r^{2}}\sin \theta \, e^{i \varphi_1} \qquad \text{and} \qquad y^{3} + i y^{4} = r\cos \theta \, e^{i \varphi_2}\,.
\end{equation*}
Which makes it clear that the supertube locus is a ring of radius $a$ in the $y^{1}-y^{2}$ plane, while the \textit{center} of the geometry is at $(r = 0, \theta = 0)$.  }:
\begin{align}
ds_4^2(\mathcal{B}) &= \frac{\Sigma}{a^{2}+r^{2}}dr^{2}+\Sigma \, d\theta^{2}+(a^{2}+r^{2})\sin^{2}\theta \, d\varphi_{1}^{2} + r^{2}\cos^{2}\theta \, d\varphi_{2}^{2}\,,
\end{align}
where:
\begin{align}
\Sigma & \equiv  r^{2}+a^{2}\cos^{2}\theta \,,
\end{align} 
and $a$ is a positive real number.

The one form $\beta$, is then taken to be the potential for a self-dual magnetic field on $\mathcal{B}$, with source along the \textit{supertube ring} ($r=0,\theta = \pi /2$):
\begin{align}
\beta = \frac{R_{y}a^{2}}{\sqrt{2}\Sigma}(\sin^{2}\theta \, d\varphi_{1} - \cos^{2}\theta \, d\varphi_{2})\,. \label{betaBody}
\end{align}

The \textit{original} single-mode superstrata were first constructed in \cite{Bena:2015bea, Bena:2016ypk, Bena:2017xbt}, and utilize the basic data: 
\begin{align}
z^{(o)}_{k,m,n} &\equiv R_{y} \frac{\Delta_{k,m,n}}{\Sigma} \cos v_{k,m,n} ~, \label{BasiczO}\\ 
\vartheta^{(o)}_{k,m,n} & \equiv -\sqrt{2} \Delta_{k,m,n} \left[ \left( (m+n)r\sin\theta+n\left( \frac{m}{k}-1 \right) \frac{\Sigma}{r\sin \theta} \right) \Omega^{(1)}\sin v_{k,m,n} \right. \notag \\
&\qquad\qquad\qquad\qquad\qquad +\left. \left( m \left(\frac{n}{k}+1 \right)\Omega^{(2)} + \left( \frac{m}{k}-1\right)n \Omega^{(3)}\right)\cos v_{k,m,n}\right] \,, \label{BasicthetaO}
\end{align}
to solve the first BPS layer. In these expressions $(\Omega^{(1)},\Omega^{(2)},\Omega^{(3)})$ are a basis of self-dual two-forms on $\mathcal{B}$, given by
\begin{align}
\Omega^{(1)} &~\equiv~ \frac{1}{(a^2 + r^2)\cos\theta} \, dr\wedge d\theta + \frac{r\sin\theta}{\Sigma} \, d\varphi_1\wedge d\varphi_2\,, \label{Omega1}\\
\Omega^{(2)} &~\equiv~  \frac{r}{a^2 + r^2} \, dr\wedge d\varphi_2 + \tan\theta\, d\theta\wedge d\varphi_1\,,\label{Omega2}\\
 \Omega^{(3)} &~\equiv~ \frac{1}{r} \, dr\wedge d\varphi_1 - \cot\theta\, d\theta\wedge d\varphi_2\,, \label{Omega3}
\end{align}
while:
\begin{align}
 \Delta_{k,m,n} & \equiv \left( \frac{a}{\sqrt{r^{2}+a^{2}}}\right)^{k} \left( \frac{r}{\sqrt{r^{2}+a^{2}}}\right)^{n}\cos^{m}\theta \sin^{k-m}\theta \,, \label{Delta1mndef}\\
v_{k,m,n} &\equiv \frac{\sqrt{2}}{R_{y}}(m+n)v+(k-m)\varphi_{1}-m \varphi_{2}\,. \label{vkmndef}
\end{align}
One must also impose constraints on the $(k,m,n)$ integers:
\begin{align}
 k,n\in \mathbb{Z}^{+} \qquad \text{and} \qquad 0\leq m \leq k\,,
\end{align}
which can be understood as required to avoid divergences coming from (\ref{Delta1mndef}).

The single-mode \textit{supercharged superstrata} were initially constructed in \cite{Ceplak:2018pws}, and solve the first BPS layer with the basic data:
\begin{align}
z^{(s)}_{k,m,n} &\equiv 0 ~, \label{BasiczS}\\ 
\vartheta^{(s)}_{k,m,n} & \equiv \sqrt{2}\Delta_{k,m,n} \left[\frac{\Sigma}{r\sin\theta} \Omega^{(1)} \sin v_{k,m,n}  + \left(\Omega^{(2)}+\Omega^{(3)} \right) \cos v_{k,m,n} \right]\,. \label{BasicthetaS}
\end{align}
for the integers $(k,m,n)$ with:
\begin{align} \label{SuperchargedModeRestriction}
1\leq m \leq k-1  \qquad \text{and} \qquad   1\leq n\,. 
\end{align}

\subsection{The full solutions}
\label{ss:Full1mnSol}
Appendix \ref{app:Constructing1mn} gives the technical details for construction of the full $(1,m,n)$ solutions, we now summarize the total solution as discussed in Section \ref{appSS:FullSol}, writing the result as concisely as possible. To do so, it is necessary to introduce two constants $(c,C_{1})$ and convenient to introduce the \textit{decoupling parameter}:
\begin{equation} \label{epsilon_def}
\epsilon_{i,j} \equiv \frac{C_{1} a^{2}}{2Q_{5}}\left(m_{i}+m_{j}+n_{i}+n_{j} \right)\,.
\end{equation}
The constant $c$ parametrizes a homogeneous solution to the second layer of the BPS equations. This constant will be fixed in (\ref{cForFlat}), as is necessary to achieve both the asymptotically AdS and flat geometries. The decoupled solution will correspond to setting the constant $C_{1} = 0$,  while flat asymptotics be given by setting $C_{1}=1$. Thus, in the decoupling limit, the decoupling parameter vanishes.

The solution to the first BPS layer is then given by\footnote{The limits of all sums have been left implicit, explicitly they are restricted by $m,m_{1},m_{1}\in \lbrace 0,1\rbrace$ and $n,n_{1},n_{2} \in \lbrace 1, 2, 3,\cdots  \rbrace$.}:
\begin{align}
Z_{1}&= C_{1}+\frac{Q_{1}}{\Sigma} + \frac{R_{y}}{2Q_{5}} \sum_{m_{1},m_{2}}\sum_{n_{1},n_{2}} \left(\frac{b_{1,m_{1},n_{1}}b_{1,m_{2},n_{2}}}{ 1+\epsilon_{1,2}} \right) z^{(o)}_{2,m_{1}+m_{2},n_{1}+n_{2}} \,, \label{Z1Body}\\
Z_{2}&= C_{1}+\frac{Q_{5}}{\Sigma} \,,\label{Z2Body} \\
Z_{4}&= \sum_{m,n}b_{1,m,n} z^{(o)}_{1,m,n}\,,
\end{align}
together with\footnote{Keep in mind that $\vartheta^{(s)}_{2,m_{1}+m_{2},n_{1}+n_{2}}$ is only well defined when $m_{1}+m_{2}=1$, the sum in (\ref{Theta2ref}) is correct if one declares that $\vartheta^{(s)}_{2,0,n_{1}+n_{2}} = \vartheta^{(s)}_{2,2,n_{1}+n_{2}} = 0 $.}:
\begin{align}
\Theta^{1}&=0 \,, \\
\Theta^{2}&= \frac{R_{y}}{2Q_{5}}\sum_{m_{1},m_{2}} \sum_{n_{1},n_{2}}  \left(\frac{b_{1,m_{1},n_{1}}b_{1,m_{2},n_{2}}}{ 1+\epsilon_{1,2}} \right) \notag \\
 & \qquad\qquad\qquad\qquad\qquad  \left( \vartheta^{(o)}_{2,m_{1}+m_{2},n_{1}+n_{2}}   -\frac{1}{2}(m_{2}-m_{1})\left(n_{2}-n_{1}\right)\vartheta^{(s)}_{2,m_{1}+m_{2},n_{1}+n_{2}}\right)  \,, \label{Theta2ref} \\
\Theta^{4}&= - 2 \sum_{m,n} b_{1,m,n}\vartheta^{(o)}_{1,m,n} \,. \label{Theta4body}
\end{align}

The second layer solution is given by:
\begin{equation}\label{Full1mnF}
\mathcal{F} =  \frac{1}{a^{2}} \left(-c^{2}+ \sum_{m} \sum_{n_{1},n_{2}} b_{1,m,n_{1}}b_{1,m,n_{2}}\Delta_{0,0,2m+n_{1}+n_{2}}\cos v_{0,0,n_{2}-n_{1}} \right)\,,
\end{equation}
and 
 \begin{al}\label{Full1mnomega}
 \omega&=\omega_0 + \frac{R_y \sin^2 \theta}{\sqrt{2} \Sigma} c^2 \, d\varphi_{1} \\
&\qquad +\frac{R_{y}}{2\sqrt{2} a^2} \sum_{m_{1},m_{2}}\sum_{n_{1},n_{2}} \Delta_{2,m_{1}+m_{2},n_{1}+n_{2}} b_{1,m_{1},n_{1}}b_{1,m_{2},n_{2}}\left( \omega^{m_{1}-m_{2}}_{n_{2}-n_{1}} - \left(\frac{\epsilon_{1,2}}{1+\epsilon_{1,2}} \right)\omega^{m_{1}+m_{2}}_{n_{1}+n_{2}} \right)\,.
 \end{al}
Where the solution for $\omega$ has been broken up into a ``subtraction mode" part:
\begin{al}
 \omega^{m_{1}-m_{2}}_{n_{2}-n_{1}}  =& \f{2(m_1-m_2)}{\sin 2\theta} d\theta \sin v_{0,m_1-m_2,n_1-n_2} \\&- \frac{1}{\Sigma}\left((2-m_1-m_2)(a^2+r^2)d\varphi_1 - (m_1+m_2)r^2d\varphi_2\right)\cos v_{0,m_1-m_2,n_1-n_2}\,,
\end{al}
and an ``addition mode" part:
\begin{al}
\omega^{m_{1}+m_{2}}_{n_{1}+n_{2}} =\omega^{(1)} + \f{n_1(m_1-1)+n_2(m_2-1)}{(n_1+n_2+m_1+m_2)}\omega^{(2)},
\end{al}
with
\begin{align}
\omega^{(1)}&= \left(\f{r}{a^2+r^2}dr - \tan\theta d\theta\right)\sin v_{2,m_1+m_2,n_1+n_2} + \frac{a^2\sin^2\theta \, d\varphi_1 - r^2 \, d\varphi_2 }{\Sigma} \cos v_{2,m_1+m_2,n_1+n_2}\,, \\
\omega^{(2)}&= -\left(\f{a^2}{r(a^2+r^2)}dr + \f{2d\theta}{\sin 2\theta}\right)\sin v_{2,m_1+m_2,n_1+n_2} - (d\varphi_1+d\varphi_2)\cos v_{2,m_1+m_2,n_1+n_2} \,. 
\end{align}

\subsection{Regularity considerations}
\label{ss:regularity}
The geometry of superstrata are a priori singular at the supertube locus, as $r \to 0$ and $\theta \to \pi /2$. This is simply illustrated by considering the $d\varphi_{1}^{2}$ component of the six dimensional metric (\ref{ds6Intro}), setting:
\begin{equation}
    r = a \epsilon \,, \qquad \theta = \frac{\pi}{2} - \epsilon \,,
\end{equation}
and expanding about $\epsilon = 0$. The result is a term of the form:
\begin{al}
\frac{1}{4 \sqrt{Q_{1}Q_{5}}\epsilon^{2}} \left( 2Q_{1}Q_{5}-(2a^{2} +c^{2})R_{y}^{2}  \right) + \mathcal{O}(\epsilon^{0})\,,
\end{al}
Thus, regularity at the supertube locus requires the fixing of a \textit{regularity condition}:
\begin{equation}
    Q_{1} Q_{5} = \frac{R_{y}^{2}}{2}(2a^{2}+c^{2})\,.
\end{equation}

Further, the ``coiffuring constraints" have already been built into the first layer BPS data (\ref{Z1Body})-(\ref{Theta4body}). As presented in Appendix \ref{app:Constructing1mn}, the most general first layer superposition would introduce the constants $(d_{k,m,n},f_{k,m,n})$ and utilize the more general:
\begin{align}
    Z_{1}&=C_{1}+\frac{Q_{1}}{\Sigma} + \frac{R_{y}}{2Q_{5}} \sum_{m_{1},m_{2}}\sum_{n_{1},n_{2}} d_{2,m_{1}+m_{2},n_{1}+n_{2}} z^{(o)}_{2,m_{1}+m_{2},n_{1}+n_{2}}\,, \\
    \Theta^{2}&=\frac{R_{y}}{2Q_{5}}\sum_{m_{1},m_{2}}\sum_{n_{1},n_{2}} \left(d_{2,m_{1}+m_{2},n_{1}+n_{2}}\vartheta^{(o)}_{2,m_{1}+m_{2},n_{1}+n_{2}} + f_{2,m_{1}+m_{2},n_{1}+n_{2}}\vartheta^{(s)}_{2,m_{1}+m_{2},n_{1}+n_{2}}  \right)  \,.
\end{align}
However, doing so would introduce singularities along the $\theta = \pi /2$ locus\footnote{For instance, one can check this is true for the $d\varphi_{1}$ component of $\omega^{(o,+)}_{m_{1}+m_{2}=0}$ presented in Appendix \ref{appSS:S120+}.}. Thus, the coiffuring constraints:
\begin{align}
    d_{2,m_{1}+m_{2},n_{1}+n_{2}} &= \frac{b_{1,m_{1},n_{1}}b_{1,m_{2},n_{2}}}{ 1+\epsilon_{1,2}} \,, \label{coiff1}\\
    f_{2,m_{1}+m_{2},n_{1}+n_{2}} &= - \frac{1}{2}(m_{2}-m_{1})(n_{2}-n_{1})\left( \frac{b_{1,m_{1},n_{1}}b_{1,m_{2},n_{2}}}{1+\epsilon_{1,2}}\right)\,, \label{coiff2}
\end{align}
have been implemented.

It was discussed in \cite{Heidmann:2019xrd, Mayerson:2020tcl}, how the decoupled solution is almost certainly regular, once the regularity and coiffuring conditions are imposed. Given the coupling to flat space provides large corrections to the decoupled solution only in the asymptotic region, where the solution approaches flat space, it follows that the flat solution is also almost certainly fully regular as well. 

\subsection{Conserved charges}
\label{ss:charges}

As will be seen in Section \ref{ss:FlatRegion}, flat asymptotics are imposed by setting:
\begin{equation}
    C_{1} = 1 \qquad \text{and} \qquad c^{2} = \sum_{m,n}b_{1,m,n}^{2}\,.
\end{equation}
Upon implementing these constraints, the conserved charges can be read off from the asymptotic behavior of the various BPS data, as outlined in Section 5.5 of \cite{Bena:2017xbt}. The following results for the charges agree exactly with those calculated using the asymptotically AdS solutions in Appendix D.5 of \cite{Mayerson:2020tcl}.

The D1 and D5 brane charges, $Q_{1}$, $Q_{5}$, are read off from the $1/r^{2}$ terms in the  expansions:
\begin{equation}
    Z_{1,2} \approx 1 + \frac{Q_{1,5}}{r^{2}}\,.
\end{equation}
From (\ref{Z1Body})-(\ref{Z2Body}), one sees that the D1, D5 charges are given by $Q_{1}$ and $Q_{5}$ respectively. 

The momentum charge, $Q_{p}$, can be read off from the expansion:
\begin{equation}
    \mathcal{F}\approx -2 Q_{p}/r^{2} + \textit{oscillatory terms}\,.
\end{equation} 
From (\ref{Full1mnF}) one sees that:
\begin{al}
    Q_{p} = \frac{1}{2}\sum_{m,n}  (m+n)(b_{1,m,n})^{2} \,.
\end{al}

The independent angular momenta, $J_{L}$, $J_{R}$, can be read off from the $d\varphi_{1} + d\varphi_{2}$ component of $\beta+\omega$ as:
\begin{al}
    \beta_{\varphi_{1}}+ \beta_{\varphi_{2}} + \omega_{\varphi_{1}} + \omega_{\varphi_{2}} \approx  \frac{\sqrt{2}}{r^{2}} \left( J_{R} - J_{L} \cos 2\theta \right) + \textit{oscillatory terms}\,.
\end{al}
From (\ref{betaBody}) and (\ref{Full1mnomega}), one then reads off that:
\begin{equation}
    J_{L} = \frac{ R_{y}}{2}a^{2}  \qquad \text{and} \qquad J_{R} = \frac{ R_{y}}{2}\left[ a^{2} +\sum_{n}(b_{1,1,n})^{2} \right] \,.
\end{equation}

\section{The decoupling limit: asymptotically AdS solutions}
\label{sect:1mnAdSSolutions}

The $(1, m, n)$ solutions in the decoupling regime were first constructed in \cite{Mayerson:2020tcl}, heavily relying on the results of \cite{Heidmann:2019xrd}. They possesses two key features, which have facilitated many calculations by supplying organizational frameworks do deal with the complexity. Namely: holomorphicity and  consistent truncation. These features are introduced in the geometric context\footnote{For gauge field considerations please refer back to \cite{Mayerson:2020tcl}.}, in this section. 

The decoupling limit is given simply by setting: 
\begin{equation}
 C_{1} \to 0 \,,  
\end{equation}
in the full solution presented in the previous section. Equivalently, since $C_1$ only appears packaged in the decoupling parameter of (\ref{epsilon_def}):
\begin{equation} 
\epsilon_{i,j} = \frac{C_{1} a^{2}}{2Q_{5}}\left(m_{i}+m_{j}+n_{i}+n_{j} \right)\,,
\end{equation}
or in $Z_{1,2}$, up to oscillating pieces as:
\begin{al}
Z_{1,2} \sim C_1 + \frac{Q_{1,5}}{\Sigma} \,,
\end{al}
one could scale:
\begin{equation}
    a^{2} \ll Q_{5} \qquad \text{and} \qquad C_1 \ll \frac{Q_{1,5}}{a^2 + r^2} \,.
\end{equation}

These limits are natural since:
\begin{itemize}
    \item $Q_{1,5}$ being large is required for the supergravity limit to be valid. 
    \item Small $a^2$ implies a long BTZ like throat, which an be inferred from the form of the geometry presented in Section \ref{ss:BTZRegion}.
    \item Considering the regime where: $r^2 \ll Q_{1,5}/C_1$, ensures one is considering the region well away from where flatness is asserted. 
\end{itemize}

\subsection{Holomorphic form}
\label{ss:HolomorphicForm}

For superstrata in the decoupling limit, it was discovered in \cite{Heidmann:2019xrd}, that the solutions can be simply expressed in terms of the complex coordinates: 
\begin{equation}
    \chi \equiv \frac{a}{\sqrt{a^{2}+r^{2}}} \sin \theta \, e^{i \varphi_{1}}\,, \qquad \mu \equiv \cot \theta \, e^{i\left( \frac{\sqrt{2}}{R_{y}}v-\varphi_{1}-\varphi_{2} \right)} \,, \qquad \xi \equiv \frac{r}{\sqrt{a^{2}+r^{2}}}e^{i \frac{\sqrt{2}}{R_{y}}v}\,,
\end{equation}
and their complex conjugates. 

These complex variables naturally appeared once the coiffuring was completed. For instance, upon defining the holomorphic function:
\begin{equation} \label{FHolo}
F(\chi, \mu, \xi) \equiv \sum_{k,m,n}b_{k,m,n}\chi^{k}\mu^{m}\xi^{n}\,,
\end{equation}
in the decoupling limit:
\begin{align}
Z_{1} &= \frac{Q_{1}}{\Sigma} + \frac{R_{y}^{2}}{4Q_{5}\Sigma}(F^{2}+\bar{F}^{2})\,, \label{Z1decoupling}\\
Z_{4} &=  \frac{R_{y}}{2\Sigma}(F+\bar{F}) \,.
\end{align}
The other BPS data and the BPS equations themselves, can similarly be written in these complex coordinates, see \cite{Heidmann:2019xrd,Mayerson:2020tcl} for details. The utility of this perspective was exploited in generating the solutions presented in \cite{Heidmann:2019xrd}, by simplifying many calculations. Utilizing the complex coordinates in this way, may be termed the \textit{holomorphic formalism}. 

The BPS data appearing in the six-dimensional metric, in the decoupling limit, can be expanded in a holomorhphic form for the $(1,m,n)$ solutions. Define:
\begin{align}\label{F01def}
    F_{0}(\xi) \equiv \sum_{n}b_{1,0,n}\xi^{n} \qquad \text{and} \qquad F_{1}(\xi) \equiv \sum_{n}b_{1,1,n}\xi^{n}\,,
\end{align}
then one can expand:
\begin{align}
    \mathcal{F}_{AdS} &= \mathcal{F}^{(p)} + c^{2} \mathcal{F}^{(c)} \,, \\
    \omega_{AdS} &= \frac{4}{\sin 2\theta} \omega_{\mu}^{(p)} \, d\theta + 2\left(\omega_{\chi}^{(0)} + \omega_{\chi}^{(p)} + c^{2} \omega_{\chi}^{(p)} \right) \, d\varphi_{1} + 2\left( \omega_{\delta}^{(0)} + \omega_{\delta}^{(p)} \right) d\varphi_{2}\,,
\end{align}
where the ``round supertube" part is given by:
\begin{equation}
    \omega_{\chi}^{(0)} = \frac{\omega_{\delta}^{(0)}}{\abs{\mu}^{2}} = \frac{R_{y}\abs{\chi}^{2}}{2\sqrt{2}(1-\abs{\chi}^{2})}\,,
\end{equation}
the ``homogeneous" part by:
\begin{equation}
    \mathcal{F}^{(c)} = -\frac{1}{a^{2}} \qquad \text{and} \qquad \omega_{\chi}^{(c)} = \frac{R_{y}\abs{\chi}^{2}}{2\sqrt{2}a^{2}(1-\abs{\chi}^{2})}\,,
\end{equation}
and the ``particular" part by:
\begin{align}
    \begin{split}
        \mathcal{F}^{(p)} &= \frac{1}{a^{2}} \left( \abs{F_{0}}^{2} +\abs{\xi}^{2}\abs{F_{1}}^{2} \right) \,, \\
        \omega_{\mu}^{(p)} &= - \frac{i R_{y}\abs{\chi}^{2}}{4\sqrt{2}a^{2}}(\mu\bar{F}_{0}F_{1} + \bar{\mu} F_{0}\bar{F}_{1}) \,,
    \end{split}
    \qquad
    \begin{split}
        \qquad \omega_{\chi}^{(p)} &= -\frac{R_{y}}{4\sqrt{2}a^{2}(1-\abs{\chi}^{2})}(\bar{\chi}\bar{F}_{0}F + \chi F_{0}\bar{F})\,, \\
        \omega_{\delta}^{(p)} &= \frac{R_{y}\abs{\xi}^{2}}{4\sqrt{2}a^{2}(1-\abs{\chi}^{2})}(\chi\mu F_{1}\bar{F} + \bar{\chi} \bar{\mu} \bar{F}_{1}F)\,.
    \end{split}
\end{align}

\subsection{Consistent truncation form}
\label{ss:CTForm}

Due to the consistent truncation identified in \cite{Mayerson:2020tcl}, the decoupled geometry can be cast in the form:
\begin{align} \label{Decoupledds6}
    ds_{6}^{2} =  \left(  \frac{\Delta}{\det m_{AB}} \right)^{1/2} ds_{3}^{2} + \frac{1}{g_{0}^{2}} \left(  \frac{\Delta}{\det m_{AB}} m_{AB}\right)^{-1/2} \mathcal{D}\mu^{A} \mathcal{D}\mu^{B}\,,
\end{align}
where the scale of the three sphere is determined by:
\begin{equation}
    g_{0} = (Q_{1}Q_{5})^{-1/4}\,.
\end{equation}
In (\ref{Decoupledds6}) we have introduced the coordinates, $\mu^{A}$, giving an embedding of the three sphere into R$^{4}$:
\begin{align}
    \mu^{1} = \sin \theta \sin \varphi_{1}\,, \qquad \mu^{2} = \sin \theta \cos \varphi_{1}\,, \qquad \mu^{3} = \cos \theta \sin \varphi_{2}\,, \qquad
      \mu^{4} = \cos \theta \cos \varphi_{2}\,,
\end{align}
the covariant derivative reads,
\begin{align}
    \mathcal{D}\mu^{A} = d\mu^{A} - 2g_{0} \tilde{A}^{AB}\mu_{B}\,,
\end{align}
and
\begin{align}
    \Delta = m_{AB} \mu^{A}  \mu^{B} \,.
\end{align}

To summarize the truncation data, it is convenient to introduce the auxiliary real functions:
\begin{equation}
    \chi_{A} = - \frac{a R_{y}g_{0}^{2}}{\sqrt{2(a^{2}+r^{2})}} \left(i F_{0}, \, F_{0}, \,  -i e^{- \frac{\sqrt{2}}{R_{y}}v}F_{1}, \,  e^{\frac{\sqrt{2}}{R_{y}}v}F_{1}  \right) +c.c. \,.
\end{equation}
One then has the 10 scalars of the symmetric:
\begin{al}
m_{AB} = \mathds{1} - \frac{1}{4} \begin{pmatrix}
\chi_{1}^{2}+\chi_{2}^{2} & 0 & \chi_{1}\chi_{3} - \chi_{2}\chi_{4} & \chi_{2}\chi_{3} + \chi_{1}\chi_{4} \\
0 & \chi_{1}^{2}+\chi_{2}^{2} & \chi_{2}\chi_{3} + \chi_{1}\chi_{4} & -\chi_{1}\chi_{3} + \chi_{2}\chi_{4} \\
\chi_{1}\chi_{3} - \chi_{2}\chi_{4} & \chi_{2}\chi_{3} + \chi_{1}\chi_{4} & \chi_{3}^{2}+\chi_{4}^{2} & 0 \\
\chi_{2}\chi_{3} + \chi_{1}\chi_{4} & -\chi_{1}\chi_{3} + \chi_{2}\chi_{4} & 0 & \chi_{3}^{2}+\chi_{4}^{2}
\end{pmatrix} \,,
\end{al}
the three dimensional external metric:
\begin{align}
    ds_{3}^{2} = \frac{R_{y}^{2}g_{0}^{2}}{2} \left[ \Omega^{2} ds_{2}^{2} - a^{2}g_{0}^{4}\left( du + dv + \frac{\sqrt{2}}{a^{2}R_{y}g_{0}^{4}}\mathscr{A}\right)^{2} \right]\,,
\end{align}
where
\begin{al}
ds_{2}^{2} = \frac{\abs{d\xi}^{2}}{\left( 1- \abs{\xi}^{2}\right)^{2}} \,, \qquad \Omega^{2} = \frac{1}{2R_{y}^{2}g_{0}^{4}}\left(4 - \chi_{A}\chi_{A} \right) \,, \qquad \mathscr{A} = \frac{i}{2} \left( \frac{\xi d\bar{\xi} - \bar{\xi} d\xi}{1- \abs{\xi}^{2}} \right)\,.
\end{al}
While the gauge fields can be expanded as:
\begin{al} \label{GaugeFieldExpansion}
\tilde{A}^{AB} = \frac{1}{\sqrt{2} a^{2}R_{y}g_{0}} \left( A_{1}\eta_{1}^{AB} + A_{2}\eta_{2}^{AB} +A_{3} \eta_{3}^{AB} + \bar{A}_{3}\bar{\eta}_{3}^{AB} \right)\,.
\end{al}

The terms appearing in the expansion of the gauge fields, (\ref{GaugeFieldExpansion}), read:
\begin{equation}
    \begin{split}
        \begin{aligned}
         A_{1} & = \frac{1}{4}\left(\chi_{1}\chi_{3}  - \chi_{2} \chi_{4}  \right) \mathbf{d} \,,\\
    A_{2} & = \frac{1}{4}\left(\chi_{1}\chi_{4}  + \chi_{2} \chi_{3}  \right) \mathbf{d} \,,
        \end{aligned}
    \end{split}
    \qquad
    \begin{split}
        \begin{aligned}
         A_{3} &= \frac{a^{2}}{2} \, dv - \frac{1}{8}(\chi_{1}^{2} + \chi_{2}^{2} - \chi_{3}^{2} - \chi_{4}^{2}) \mathbf{d}\,, \\
    \bar{A}_{3}& = - \left( \frac{a^{2}+2r^{2}}{2} \right) dv + \left( 1 + \frac{1}{8}\chi_{A}\chi_{A} \right)\mathbf{d}\,,
        \end{aligned}
    \end{split}
\end{equation}
where
\begin{equation}
    \mathbf{d} = \frac{1}{\Omega^{2}} \left[ a^{4}(du + dv) + \frac{2r^{2}}{R_{y}^{2}g_{0}^{4}}dv \right]\,,
\end{equation}
and the anti-symmetric $4 \times 4$ `t Hooft matrices have been introduced:
\begin{align}
    \eta_{1}^{AB} &= \begin{pmatrix} 0 & \sigma_{x} \\ -\sigma_{x} & 0 \end{pmatrix}\,, \qquad \eta_{2}^{AB} = \begin{pmatrix} 0 & -\sigma_{z} \\ \sigma_{z} & 0 \end{pmatrix} \,, \qquad
    \eta_{3}^{AB} = \begin{pmatrix} i\sigma_{y} & 0 \\ 0 & i\sigma_{y} \end{pmatrix} \,, \\
    \bar{\eta}_{1}^{AB} &= \begin{pmatrix} 0 & -i\sigma_{y} \\ i\sigma_{y} & 0 \end{pmatrix}\,, \qquad \bar{\eta}_{2}^{AB} = \begin{pmatrix} 0 & -\mathbb{1} \\ \mathbb{1} & 0 \end{pmatrix} \,, \qquad
    \bar{\eta}_{3}^{AB} = \begin{pmatrix} i\sigma_{y} & 0 \\ 0 & -i\sigma_{y} \end{pmatrix} \,.
\end{align}

\section{The geometry}
\label{sect:TheGeometry}

This section elucidates the geometry of the $(1, m, n)$ asymptotically flat superstrata, by expanding it in three limits: 
\begin{itemize}
    \item The \textit{asymptotic region}, $r \to \infty$, where the geometry becomes flat. 
    \item The \textit{supertube cap region}, $r \to 0$ and $\theta = \pi/2$, where the geometry caps off smoothly, just above where the horizon of the $a\to 0 $ black hole would sit. 
    \item The \textit{long BTZ throat region}, where the geometry approximates a BTZ black hole fibered with a deformed three sphere.  
\end{itemize}

\subsection{Asymptotically flat region}
\label{ss:FlatRegion}
To reach the asymptotically flat region, first set $C_1 = 1$ in the solution presented in Section \ref{ss:Full1mnSol}, then expand the metric (\ref{ds6Intro}), as $r\to \infty$. To leading order this gives: 
\begin{equation} \label{ds6FlatExpansion1}
    \begin{aligned}
    \lim_{r\to \infty}ds_{6}^{2} & =    \left(\frac{c^{2} - \abs{F_{0}^{(\infty)}}^{2} - \abs{F_{1}^{(\infty)}}^{2}}{a^{2}} \right)  dv^{2} -2 \, du \, dv  +dr^2 +r^{2} d\Omega_{3}^{2}\,.
    \end{aligned}
\end{equation}
Where $d\Omega_{3}^{2}$ is the canonical metric on S$^3$:
\begin{equation}
    d\Omega_{3}^{2} = d\theta^{2}  + \sin^2 \theta \, d\varphi_{1}^{2} + \cos^2 \theta \, d\varphi_{2}^{2}\,\,,
\end{equation}
and utilizing (\ref{F01def}):
\begin{align}
    F_{0,1}^{(\infty)}(v) \equiv \lim_{r \to \infty}F_{0,1}(v,r) = \lim_{\abs{\xi}\to 1}F_{0,1}(\xi)\,.
\end{align}

The metric (\ref{ds6FlatExpansion1}) is almost flat space, the discrepancy being due to the $dv^2$ term. However, this can be rectified by setting\footnote{We now see why it was prudent to add the homogeneous part to the solution in Section \ref{ss:Full1mnSol}. It was required to allow for the correct asymptotics. }:
\begin{equation} \label{cForFlat}
    c^2 = \sum_{m,n} b_{1,m,n}^{2}\,,
\end{equation}
in order to remove the non-oscillatory RMS part of the $dv^{2}$ term. The oscillatory part can then be removed by a gauge transformation of the form \eqref{gaugetrf}, with\footnote{It is common in the superstrata literature to use a dot to denote a derivative with respect to $v$, a convention we utilize here.}: 
\begin{align}
    \dot{U} =  \frac{1}{2a^{2}}\left( c^{2} - \abs{F_{0}^{(\infty)}}^{2} - \abs{F_{1}^{(\infty)}}^{2} \right)\,.
\end{align}
Then, utilizing the $(t,y)$ coordinates implicitly defined by (\ref{uvDefUlt}), the asymptotic metric takes the canonically flat form:
\begin{equation}
    \lim_{r\to \infty}ds_{6}^{2} =  -dt^2 + dy^2 + dr^{2}+  r^{2} d\Omega_{3}^{2}\,.
\end{equation}

\subsection{Supertube cap region}
\label{ss:CapRegion}
To illuminate the smooth capping off of the solution at the supertube locus, we introduce the dimensionless radial variable:
\begin{equation}
    \rho = r /a\,,
\end{equation}
set $\theta = \pi/2$ and expand about $\rho = 0$. We present this expansion with the sphere contributions factored out as\footnote{We will not dwell on the form of the fields $\tensor{\mathcal{A}}{^{i}_{j}}$.}:
\begin{al}
ds^{2}_{6} = ds_{3}^{2} + \frac{1}{2}\sum_{x^{i} \in \lbrace  \theta, \varphi_{1}, \varphi_{2}\rbrace}\sum_{x^{j} \in \lbrace  t,y,\rho\rbrace }  \left(dx^{i} + \tensor{\mathcal{A}}{^{i}_{j}} dx^{j} \right)^{2}\,.
\end{al}

The result is:
\begin{al} \label{supertubelocus}
ds_{3}^{2} &= \sqrt{Q_{1}Q_{5}}\left[ \frac{d\rho^2}{1-\rho^2} - \frac{a^{4}R_{y}^{2}}{Q_{1}^{2}Q_{5}^{2}}(1+\rho^2)dt^{2} + \frac{\rho^{2}}{R_{y}^{2}}\left(dy +\frac{c^{2}R_{y}^{2}}{2Q_{1}Q_{5}}dt  \right)^{2} \right] \\ 
& \qquad +\frac{a^{6}R_{y}^{2}C_{1}}{(Q_{1}Q_{5})^{3/2}}\left( \frac{Q_{1}+Q_{5}}{Q_{1}Q_{5}+a^{2}C_{1
}(Q_{1}+Q_{5})}\right) dt^2  \\
&\qquad\qquad  -\frac{g_{0}^2}{2}\left(\abs{F_{0}}^{2} + \abs{F_{1}}^{2}\right) \left[ \rho^{2}(dt + dy)^{2}+ \frac{R_{y}^{2}d\rho^{2}}{(1+\rho^{2})^{2}} \right]+\mathcal{O}(\rho^2)  \,.
\end{al}
Where all the terms in the third line are $\mathcal{O}(\rho^{2})$, the corrections that have not been written are quadratic in the $b_{k,m,n}$ and vanish when $C_{1}\to 0$. The key takeaway from (\ref{supertubelocus}) is that the flat corrections, as captured by the second line, only alters the scaling of $t$, whilst leaving the radius of the $y$-circle unchanged as $\rho \to 0$. Thus the smooth capping off, in an AdS$_{3}$ like manner at the supertube locus, given by the first line in (\ref{supertubelocus}), is unaltered by the flat space coupling. 

A priori, the flat space coupling may also destroy regularity within the supertube ring, for $\rho = 0$ and $\theta < \pi/2$. Since the superstrata corrections in line three of (\ref{supertubelocus}) are subleading as $\rho \to 0$, the regularity analysis inside the ring is identical to that of the round supertube coupled to flat space. A demonstration of the regularity of the round supertube can be found in \cite{Lunin:2002iz,Mathur:2005zp}. It is interesting to note that, in the decoupled solution the circle that caps off is universal to the ring and its interior, whilst for the asymptotically flat solution it is a non-trivial combinations of the periodic directions.

\subsection{Long BTZ throat region}
\label{ss:BTZRegion}
To demonstrate the region in which the geometry approximates a long throated BTZ black hole, fibred with a three sphere, we first introduce the dimensionless coordinates:
\begin{equation}
(\tilde{t},\tilde{y},\tilde{r} ) = (Q_{1}Q_{5})^{-1/4} (t,y,r)= g_{0} (t,y,r)\,.
\end{equation}
Then we expand the metric into the form:
\begin{al} \label{ds6BTZexpansion}
   ds_{6}^{2} &=  \left(  \frac{\Delta}{\det m_{AB}} \right)^{1/2} ds_{3}^{2} +\frac{1}{g_{0}^{2}} \left(  \frac{\Delta}{\det m_{AB}} \right)^{-1/2} \mathcal{D}\mu^{A} m_{AB} \mathcal{D}\mu^{B}  \\
   & \qquad \qquad \qquad \qquad  +   ds^{2}_{3,C_{1}}  + \frac{1}{2}\sum_{x^{i} \in \lbrace  \theta, \varphi_{1}, \varphi_{2}\rbrace}\sum_{x^{j} \in \lbrace  \tilde{t},\tilde{y},\tilde{r}\rbrace }  \left(dx^{i} + \tensor{\mathcal{A}}{^{i}_{j}} dx^{j} \right)^{2}\,.
\end{al}
The first line of (\ref{ds6BTZexpansion}) corresponds to the decoupling limit contributions, given by setting\footnote{I.e. the asymptotically AdS decoupled solution, presented in the form of the consistent truncation, as given in Section \ref{ss:CTForm}.} $C_{1} =0$. While, the second line of (\ref{ds6BTZexpansion}) gives the ``flat corrections," involving terms that vanish when $C_{1} \to 0$, that are not captured by the consistent truncation ansatz. We will neglect the terms, mixing the $(d\tilde{t},d\tilde{y},d\tilde{r})$ and $(d\theta, d\varphi_1, d\varphi_2)$, though one can check that these terms are perfectly well behaved and don't materially alter the following story. To present the results cleanly, it will be useful to define:
\begin{al}
\tilde{F}^{m}_{n_{1},n_{2}} &\equiv \frac{1}{2}\sum_{n_{1},n_{2}} b_{1,m,n_{1}}b_{1,m,n_{2}}(n_{1}+n_{2}) e^{\frac{i\sqrt{2}}{R_{y}}(n_{2}-n_{1})v}\,.
\end{al}

Working to first order in $a$, we find that the warp factors expand as:
\begin{align}
    \left(  \frac{\Delta}{\det m_{AB}} \right)^{\pm 1/2} &= 1 + \mathcal{O}(a^2) \,, 
\end{align}
the truncated metric expands as:

\begin{align}
    ds_{3}^{2} & = \sqrt{Q_{1}Q_{5}} \left( \frac{d\tilde{r}^{2}}{\tilde{r}^{2}} - \tilde{r}^{2}\, d\tilde{t}^{2} + \tilde{r}^{2} \, d\tilde{y}^{2} \right) \notag\\
    & \qquad + \sqrt{Q_{1}Q_{5}}\left\lbrace \frac{\tilde{F}^{0}_{n_{1}+n_{2}} + \tilde{F}^{1}_{n_{1}+n_{2}}}{\sqrt{2} R_{y}c} -\frac{\sqrt{Q_{1} Q_{5}}}{4R_{y}^{2} c^{4}} \left( \abs{F_{0}^{(\infty)}}^{2} + \abs{F_{1}^{(\infty)}}^{2} - c^{2}\right)^{2} \right\rbrace  \left(d\tilde{t}+ d\tilde{y} \right)^{2} + \mathcal{O}(a^2) \,,
\end{align}
while the flat corrections expand as:
\begin{equation} \label{ds3C1Body}
        ds^{2}_{3,C_{1}} =   P_{1} (\tilde{r})\left\lbrace X(v,\tilde{r},\theta,\varphi_{1},\varphi_{2}) \left[(d\tilde{t}+d\tilde{y})^2 - 2\tilde{r}^{2} (d\tilde{t} + d\tilde{y})d\tilde{y} \right] + \left(\frac{1}{1-g_{0}^{2}P_{1}(\tilde r)} \right) \frac{d\tilde{r}^{2}}{\tilde{r}^{2}} 
 \right\rbrace +\mathcal{O}(a^2)\,,
\end{equation}
where 
\begin{al}
P_{1}(\tilde{r}) =  \frac{1}{g_{0}^{2}}\left( 1 - \frac{1}{\sqrt{(1+ C_{1}g_{0}^{2}Q_{1}\tilde{r}^{2})(1+ C_{1}g_{0}^{2}Q_{5}\tilde{r}^{2})}} \right) \,,
\end{al}
$X(v,\tilde{r},\theta,\varphi_{1},\varphi_{2})$ is of $\mathcal{O}(b_{k,m,n}^{2})$ and appears to have no convenient closed form or expansion. 

To give a sense of the complexity of $X(v,\tilde{r},\theta,\varphi_{1},\varphi_{2})$, we include its explicit form, including only modes with $n \in \lbrace 1,2,3 \rbrace$, whence it can be expanded as:
\begin{al}
X(v,\tilde{r},\theta,\varphi_{1},\varphi_{2}) = \tilde{r}^{2} - \frac{1}{2} g_{0}^{2} X_{1} + \frac{R_{y}^{2}g_{0}^{6}}{4}\left( P_{1}g_{0}^{2}-1 \right)\left( X_{2}^{2} + X_{3}^{2} + X_{4}^{2} \right)\,,
\end{al}
where:
\begin{al}
    X_{1} &=   b_{1,0,1}^{2} + 2 b_{1,0,2}^{2}+3 b_{1,0,3}^{2}+2 b_{1,1,1}^{2}+3b_{1,1,2}^{3}+4 b_{1,1,3}^{3}    \\
& \qquad + \left[ b_{1,0,2}(3b_{1,0,1}+5b_{1,0,3})+b_{1,1,2}(5b_{1,1,1}+7b_{1,1,3}) \right] \cos v_{0,0,1} \\
& \qquad \qquad + 2\left(2b_{1,0,1}b_{1,0,3}+3b_{1,1,1}b_{1,1,3} \right) \cos v_{0,0,2} \,,
\end{al}
and
\begin{al}
X_{2} &= \left( b_{1,0,1}b_{1,1,1} +b_{1,0,2}b_{1,1,2} + b_{1,0,3}b_{1,1,3} \right) \cos v_{ 0,1,0} + \left( b_{1,0,1}b_{1,1,2} + b_{1,0,2}b_{1,1,3} \right)\cos v_{0,1,1}  \\
& \qquad + b_{1,0,1}b_{1,1,3}\cos v_{0,1,2}  +\left( b_{1,0,2}b_{1,1,1}+b_{1,0,3}b_{1,1,2} \right)\cos v_{0,1,-1} +b_{1,0,3}b_{1,1,1}\cos v_{0,1,-2} \,, \\
X_{3} &= b_{1,1,1}^{2} +b_{1,1,2}^{2}+b_{1,1,3}^{2} - \left( b_{1,0,2}b_{1,0,1}+b_{1,0,2}b_{1,0,3}-b_{1,1,1}b_{1,1,2} - b_{1,1,2}b_{1,1,3} \right)\cos v_{0,0,1}  \\
    & \qquad +\left(b_{1,1,1}b_{1,1,3} - b_{1,0,1}b_{1,0,3}  \right)\cos v_{0,0,2} \,,\\
X_{4} &= \left( b_{1,0,1}b_{1,1,1} +b_{1,0,2}b_{1,1,2} + b_{1,0,3}b_{1,1,3} \right) \sin v_{ 0,1,0} + \left( b_{1,0,1}b_{1,1,2} + b_{1,0,2}b_{1,1,3} \right)\sin v_{0,1,1} \\
& \qquad + b_{1,0,1}b_{1,1,3}\sin v_{0,1,2}  +\left( b_{1,0,2}b_{1,1,1}+b_{1,0,3}b_{1,1,2} \right)\sin v_{0,1,-1} +b_{1,0,3}b_{1,1,1}\sin v_{0,1,-2} \,.
\end{al}

The standard BTZ metric in our coordinates, for the patch outside the horizon, takes the form:
\begin{equation}
    ds_{BTZ}^{2} = \sqrt{Q_{1}Q_{5}} \left( \frac{d\tilde{r}^{2}}{\tilde{r}^{2}} - \tilde{r}^{2}\, d\tilde{t}^{2} + \tilde{r}^{2} \, d\tilde{y}^{2} \right) + \alpha(d\tilde{y}+d\tilde{t})^{2}\,,
\end{equation}
where $\alpha$ is a constant upon which the charges depend. Thus, one sees that \textit{external} part of the six-dimensional metric:
\begin{equation} \label{dsexternal}
ds_{3,\text{external}}^{2} =\left(  \frac{\Delta}{\det m_{AB}} \right)^{1/2} ds_{3}^{2} +   ds^{2}_{3,C_{1}}\,,
\end{equation}
is the BTZ metric with two types of corrections:
\begin{itemize}
    \item $\mathcal{O}(b_{k,m,n}^{2})$ corrections that are oscillatory in $v$.
    \item Corrections coming from the flat space coupling, that are suppressed by a factor of $P_{1}$, which ensures these terms vanish in the decoupling limit. 
\end{itemize}

\section{Discussion and conclusions}
\label{sect:Conclusions}

Having constructed the full $(1,m,n)$ superstrata in Section \ref{sect:full1mn}, considered the decoupled AdS limit in Section \ref{sect:1mnAdSSolutions} and illustrated various limits of the asymptotically flat geometries in Section \ref{sect:TheGeometry}, it is now worth revisiting the two main research questions of this paper: 
 \begin{itemize}
     \item \textit{What features of microstate geometries, in the decoupling limit, are altered or retained when more realistic asymptotics are imposed?}
     \item \textit{To what extent are the currently accepted results or intuition, in the microstate geometry program, artefacts of working in the decoupling limit?}
 \end{itemize}

To first order, an answer comes from comparing $(\mathcal{F},\omega)$, key data in the metric expansion (\ref{ds6Intro}), between the full solution (\ref{Full1mnF})-(\ref{Full1mnomega}) and the decoupled solution, given by setting $\epsilon_{1,2}=0$. One sees immediately that the ``addition modes" in $\omega$, make the full solution's geometry vastly more involved than the decoupled solution's geometry, especially when many more modes are considered in the superposition. 

To second, the explicit implications this additional geometric structure should be explored. Some straightforward implications demonstrated in this paper include:
\begin{itemize}
    \item The full solution is no longer simply written in terms of the holomorphic formalism.
    \item The full solution does not belong to the consistent truncation to which the  decoupled solution does.
    \item Changing the asymptotics has effects on the geometry, even deep within the long throat.  
\end{itemize}
Beyond these straightforward observations, it is now possible to revisit many of the calculations previously performed in the decoupling limit, to determine the corrections due to the flat space coupling. We leave such intriguing analysis to future projects. 

\subsection{Holomorphic considerations}

The holomorphic formalism traces its origins to the coiffuring relations, required to ensure regularity of the decoupled solution. However, coupling the solutions to flat space alters the coiffuring relations, see (\ref{coiff1})-(\ref{coiff2}). So it is not obvious that the holomorphic formalism survives the transition. The simplest way to probe this is by considering (\ref{Z1Body}):
\begin{equation} \label{Z1conclusion}
    Z_{1}= C_{1}+\frac{Q_{1}}{\Sigma} + \frac{R_{y}}{2Q_{5}} \sum_{m_{1},m_{2}}\sum_{n_{1},n_{2}} \left(\frac{b_{1,m_{1},n_{1}}b_{1,m_{2},n_{2}}}{ 1+\epsilon_{1,2}} \right) z^{(o)}_{2,m_{1}+m_{2},n_{1}+n_{2}}\,,
\end{equation}
where (\ref{epsilon_def}):
\begin{equation}
\epsilon_{i,j} = \frac{C_{1} a^{2}}{2Q_{5}}\left(m_{i}+m_{j}+n_{i}+n_{j} \right)\,.
\end{equation}
When trying to reformulate this in the complex variables, one notes that:
\begin{equation}
    z^{(o)}_{2,m,n} = \chi^{2}\mu^{m}\xi^{n} + \bar{\chi}^{2}\bar{\mu}^{m}\bar{\xi}^{n}\,,
\end{equation}
but then the $\epsilon_{1,2}$ in (\ref{Z1conclusion}) becomes very cumbersome. This can not be generated cleanly from some holomorphic functions.

On the other hand, in the decoupling limit, $Z_{1}$ does has the very simple holomorphic form (\ref{Z1decoupling}):
\begin{equation}
    Z_{1} = \frac{Q_{1}}{\Sigma} + \frac{R_{y}^{2}}{4Q_{5}\Sigma}(F^{2}+\bar{F}^{2})\,,
\end{equation}
with (\ref{FHolo}):
\begin{equation}
F(\chi, \mu, \xi) \equiv \sum_{k=1}^{1}\sum_{m=0}^{1}\sum_{n}b_{k,m,n}\chi^{k}\mu^{m}\xi^{n}\,.
\end{equation}
Thus, it appears that the utility of the holomorphic formalism identified in \cite{Heidmann:2019xrd}, was a consequence of working in the decoupling regime. Rather than reflecting an underlying structure, such as supersymmetry.

\subsection{Consistent truncation considerations}
From the expansion given of the asymptotically flat $(1,m,n)$ solution in the long BTZ throat region of Section \ref{ss:BTZRegion}, it is clear that the full solution can not belong to a consistent truncation on a three sphere. To see this, consider the external three dimensional part of the metric (\ref{dsexternal}):
\begin{align}
   ds_{3,\text{external}}^{2} \equiv  \left(  \frac{\Delta}{\det m_{AB}} \right)^{1/2} ds_{3}^{2}   +   ds^{2}_{3,C_{1}}\,.
\end{align}
Considering the form of $ds_{3,C_{1}}^{2}$ in (\ref{ds3C1Body}), it is clear that the dependence on the spherical coordinates,  $(\theta,\varphi_{1},\varphi_{2})$, can not be factored out into a simple ``warp factor." Thus the dynamics of the sphere can not couple in the requisite way, to the external part of the geometry, for a consistent truncation to be valid. 

Although this removes one of the most powerful supergravity tools for simplifying analysis, it also opens a window into uncovering new physics. For instance, it shows that the consistent truncation is missing some supergravity degrees of freedom, necessary to couple to flat space. Can these degrees of freedom be identified in some convenient manner? It is conceivable that the full $(1,m,n)$ solution presented herein could be used to answer such a question.  

\subsection{UV/IR mixing considerations}
An intriguing feature of quantum gravity is the possibility for subtle UV/IR mixing effects. Wherein the short distance, UV physics, may qualitatively alter the long distance, IR physics. A result that runs counter to the effectiveness of re-normalization in quantum field theories, where the UV degrees of freedom can be discarded. 

In the context of microstate geometries, the UV region corresponds to the asymptotic region, while the IR region corresponds to the region deep down the long throat, including the cap. With this in mind, consider the conceptual processes of producing asymptotically flat superstrata:
\begin{enumerate}
    \item Change UV structure, from AdS to flat asymptotics.
    \item Regularity of superstrata requires altered coiffuring relations.
    \item Altered coiffuring leads to new structure deep down the long throat and at the cap.
\end{enumerate}
Thus, the additional structure, relative to the decoupled solutions, presented in Sections \ref{ss:CapRegion} and \ref{ss:BTZRegion}. Might be suitably interpreted as artifacts of UV/IR mixing in the microstate geometry program.

\section*{Acknowledgments}
\vspace{-2mm}
We are  grateful to Daniel Mayerson and Pierre Heidmann for insightful discussions on this project. Further, the work of RAW was supported in part by the Research Foundation - Flanders (FWO). 
\vspace{1cm}
\appendix

\section{Constructing the solutions}
\label{app:Constructing1mn}
This appendix gives details on how the full $(1,m,n)$ asymptotically flat solutions, of Section \ref{ss:Full1mnSol}, were generated. Schematically it involves:
\begin{itemize}
    \item Developing an ansatz for the first BPS layer data, with enough modes to allow for regularity of the final solution.
    \item Decomposing the second layer sources into ``basic" parts.
    \item Identifying solutions to the basic second layer sources that already appear in the literature.
    \item Solving the second layer for the basic sources that don't already appear in the literature. 
    \item Utilizing the semi-linear property of the BPS equations to superpose the basic solutions.
    \item Tuning the constants in the first layer ansatz such that the full solution is regular, a process known as \textit{coiffuring}.
\end{itemize}
In practice, for some basic sources, the last three points can only be completed simultaneously. This is in contrast to the solutions purely in the decoupled limit, where each step can be completed independently, explicitly. 
 
\subsection{First layer ansatz}
\label{appSS:AppSub1}
When searching for new multi-mode superstrata, one begins by specifying the modes in $Z_4$, that one would like to consider. In this paper we consider all possible modes of the form $(1,m,n)$. Since $0 \leq m \leq k = 1$ for the original superstrata modes and there are no supercharged modes of this form, this leaves:
\begin{equation}
Z_{4}= \sum_{m=0}^{1}\sum_{n=1}^{\infty} b_{1,m,n} z^{(o)}_{1,m,n}\,.
\end{equation}
In this expansion, the constants $b_{1,m,n}$ have been introduced, these will be moduli of the final solution. In the remainder of this appendix. the limits of sums over $m$ and $n$ indices will be left implicit.

After specifying $Z_4$, an ansatz for $Z_2$ is then postulated that involves all the ``quadratic" modes of those appearing in $Z_{1}$, with arbitrary coefficients. These quadratic modes will be vital in the coiffuring process required for regularity, allowing partial cancellation of quadratic terms coming from $Z_{4}Z_{4}$ and $Z_{4}\Theta^{4}$ like terms in the second BPS layer sources. Meanwhile, the form of $Z_{2}$ is unchanged for all superstrata solutions, thus we postulate a full ansatz for the $Z_{I}$ of:
\begin{equation} \label{ZAnsatzFull}
\begin{aligned}
Z_{1}&=C_{1}+\frac{Q_{1}}{\Sigma} + \frac{R_{y}}{2Q_{5}} \sum_{m_{1},m_{2}}\sum_{n_{1},n_{2}} d_{2,m_{1}+m_{2},n_{1}+n_{2}} z^{(o)}_{2,m_{1}+m_{2},n_{1}+n_{2}}\,, \\
Z_{2}&= C_{1}+\frac{Q_{5}}{\Sigma} \,, \\
Z_{4}&= \sum_{m,n} b_{1,m,n} z^{(o)}_{1,m,n}  \,.
\end{aligned}
\end{equation}

The corresponding ansatz for the $\Theta^{I}$ follows directly from the basic superstrata solutions (\ref{BasiczO})-(\ref{BasicthetaO}) and (\ref{BasiczS})-(\ref{BasicthetaS}), so are given by\footnote{Given the mode restriction in (\ref{SuperchargedModeRestriction}), in these sums one must declare that $\vartheta^{(s)}_{2,0,n_{1}+n_{2}} = \vartheta^{(s)}_{2,2,n_{1}+n_{2}} = 0 $. }:
\begin{equation} \label{ThetaAnsatzFull}
\begin{aligned}
\Theta^{1}&=0 \,, \\
\Theta^{2}&=\frac{R_{y}}{2Q_{5}}\sum_{m_{1},m_{2}}\sum_{n_{1},n_{2}} \left(d_{2,m_{1}+m_{2},n_{1}+n_{2}}\vartheta^{(o)}_{2,m_{1}+m_{2},n_{1}+n_{2}} + f_{2,m_{1}+m_{2},n_{1}+n_{2}}\vartheta^{(s)}_{2,m_{1}+m_{2},n_{1}+n_{2}}  \right)  \,, \\
\Theta^{4}&=-2 \sum_{m,n} b_{1,m,n}\vartheta^{(o)}_{1,m,n}\,.
\end{aligned}
\end{equation}

The task now is to solve the second BPS layer, given this first layer ansatz, while restricting the coefficients $d_{2,m_{1}+m_{2},n_{1}+n_{2}}$ and $f_{2,m_{1}+m_{2},n_{1}+n_{2}}$ to ensure regularity. This restriction of coefficients is termed ``coiffuring."

\subsection{Second BPS layer source decomposition}
\label{appSS:AppSub2}


Given the first layer ansatz (\ref{ZAnsatzFull})-(\ref{ThetaAnsatzFull}), the sources can be written in terms of a double sum as:
\begin{align} \label{SourceDecomp}
\mathcal{S}_{1,2} &= \sum_{m_{1},m_{2}}\sum_{n_{1},n_{2}} \left(\mathcal{S}_{1,2}^{(-)}+\mathcal{S}_{1,2}^{(o,+)} + \mathcal{S}_{1,2}^{(s,+)} \right) \,.
\end{align}
Where the \textit{basic sources}: $\mathcal{S}_{1,2}^{(-)}$, $\mathcal{S}_{1,2}^{(o,+)}$ and $\mathcal{S}_{1,2}^{(s,+)}$, have been introduced. These basic sources depend implicitly on the mode numbers $(m_{1},m_{2},n_{1},n_{2})$.

The basic sources correspond to; the the ``addition mode" terms coming from the original superstrata contributions in the ansatz:
\begin{align*}
\mathcal{S}_{1}^{(o,+)} &=   \frac{R_{y}}{2}\left[\left(\frac{C_{1}}{ Q_{5}}+ \frac{1}{\Sigma} \right)d_{2,m_{1}+m_{2},n_{1}+n_{2}}  -\frac{1}{ \Sigma}   b_{1,m_{1},n_{1}}b_{1,m_{2},n_{2}} \right] \vartheta^{(o)}_{2,m_{1}+m_{2},n_{1}+n_{2}} \,, \\
\mathcal{S}_{2}^{(o,+)} &=  - \frac{1}{R_{y}} (m_{1}+m_{2}+n_{1}+n_{2})^{2} \left[\left(\frac{C_{1}}{Q_{5}}+ \frac{1}{\Sigma} \right)d_{2,m_{1}+m_{2},n_{1}+n_{2}}  -\frac{1}{ \Sigma}   b_{1,m_{1},n_{1}}b_{1,m_{2},n_{2}} \right] z^{(o)}_{2,m_{1}+m_{2},n_{1}+n_{2}}\,,
\end{align*}
the ``addition mode" terms coming from the supercharged contributions in the ansatz:
\begin{align*}
\mathcal{S}_{1}^{(s,+)} & = \frac{R_{y}}{2}\left[\left(\frac{C_{1}}{ Q_{5}}+\frac{1}{\Sigma} \right)f_{2,m_{1}+m_{2},n_{1}+n_{2}}  +\frac{1}{ 2\Sigma}   (n_{2}-n_{1})(m_{2}-m_{1}) b_{1,m_{1},n_{1}}b_{1,m_{2},n_{2}} \right] \vartheta^{(s)}_{2,m_{1}+m_{2},n_{1}+n_{2}}  \,, \\
\mathcal{S}_{2}^{(s,+)} & = 0 \,,
\end{align*}
and the ``subtraction mode" terms:
\begin{align*}
\mathcal{S}_{1}^{(-)} & = \f{R}{\sqrt{2} \Sigma} \Delta_{2,m_1+m_2,n_1+n_2} b_{1,m_1,n_1}b_{1,m_2,n_2}   \\
& \qquad \left\lbrace \left[(m_1-m_2+n_1-n_2)r\sin\theta\vphantom{\f{R}{2}}+ ((m_1-1)n_1 - (m_2-1)n_2)\f{\Sigma}{r\sin\theta}\right]\Omega^{(1)}\sin v_{0,m_1-m_2,n_1-n_2} \right.\\
& \qquad\qquad\left.+ \left[(m_1(1+n_1)+m_2(1+n_2))\Omega^{(2)} + ((m_1-1)n_1 + (m_2-1)n_2)\Omega^{(3)}\right]\cos v_{0,m_1-m_2,n_1-n_2}\right\rbrace \,, \\
\mathcal{S}_{2}^{(-)} & =  \frac{1}{R_{y}}  \Delta_{2,2m_{1},2n_{1}}b_{1,m_{1},n_{1}}b_{1,m_{2},n_{2}} \\
& \qquad \left\lbrace \frac{1}{\Sigma}(m_{2}-m_{1}+n_{2}-n_{1})^{2} + 2\left[\frac{(1-m_{1})(1-m_{2})n_{1}n_{2}}{r^{2}\sin^{2}\theta} + \frac{m_{1}m_{2}(1+n_{1})(1+n_{2})}{(a^{2}+r^{2})\cos^{2}\theta} \right]    \right\rbrace  z^{(o)}_{0,m_{2}-m_{1},n_{2}-n_{1}} \,.
\end{align*}

\subsection{$S_{1,2}^{(o,+)}$ solutions}
\label{appSS:S120+}
When solving the second layer with the ``addition mode" original basic source:
\begin{align*}
\mathcal{S}_{1}^{(o,+)} &=   \frac{R_{y}}{2}\left[\left(\frac{C_{1}}{ Q_{5}}+ \frac{1}{\Sigma} \right)d_{2,m_{1}+m_{2},n_{1}+n_{2}}  -\frac{1}{ \Sigma}   b_{1,m_{1},n_{1}}b_{1,m_{2},n_{2}} \right] \vartheta^{(o)}_{2,m_{1}+m_{2},n_{1}+n_{2}} \,, \\
\mathcal{S}_{2}^{(o,+)} &=  - \frac{1}{R_{y}} (m_{1}+m_{2}+n_{1}+n_{2})^{2} \left[\left(\frac{C_{1}}{Q_{5}}+ \frac{1}{\Sigma} \right)d_{2,m_{1}+m_{2},n_{1}+n_{2}}  -\frac{1}{ \Sigma}   b_{1,m_{1},n_{1}}b_{1,m_{2},n_{2}} \right] z^{(o)}_{2,m_{1}+m_{2},n_{1}+n_{2}}\,,
\end{align*}
it is convenient to work with the instances $m_{1}+m_{2} = 0,1,2$ separately. In each instance the solution can be adapted from those presented in \cite{Bena:2017xbt}.

When $m_{1} + m_{2} = 0$, the solution is given by:
\begin{align*}
    \mathcal{F}_{m_{1}+m_{1}=0}^{(o,+)} &= 0 \,, \\
    \mathcal{\omega}_{m_{1}+m_{1}=0}^{(o,+)} &= -\frac{C_{1}R_{y}}{4\sqrt{2}Q_{5}}(n_{1}+n_{2}) d_{2,0,n_{1}+n_{2}}\Delta_{2,0,n_{1}+n_{2}}  \\
    & \qquad\qquad\qquad\left[ \sin v_{2,0,n_{1}+n_{2}}  \left(\frac{dr}{r} - \tan \theta \, d\theta  \right)  + \cos v_{2,0,n_{1}+n_{2}}\, d\varphi_{2} \right]  \\
    & \qquad + \frac{R_{y}}{2\sqrt{2}}\left(d_{2,0,n_{1}+n_{2}}-b_{1,0,n_{1}}b_{1,0,n_{2}} \right)\Delta_{2,0,n_{1}+n_{2}}\\
    & \qquad \qquad\qquad\qquad\qquad \left[ \frac{2}{a^{2}\sin 2\theta}\sin v_{2,0,n_{1}+n_{2}} d\theta + \frac{1}{\Sigma} \cos v_{2,0,n_{1}+n_{2}}\left(d\varphi_{1} + \frac{r^{2}}{a^{2}}(d\varphi_{1}-d\varphi_{2})\right)  \right]\,.
\end{align*}
When $m_{1} + m_{2} = 1$, the solution is known when the coiffuring condition:
\begin{equation} \label{CoifdApp1}
    d_{2,m_{1}+m_{2},n_{1}+n_{2}} = \frac{b_{1,m_{1},n_{1}}b_{1,m_{2},n_{2}}}{ 1+\epsilon_{1,2}}\,, 
\end{equation}
is imposed. With this coiffuring constraint:
\begin{align*}
\mathcal{F}^{(o,+)}_{m_{1} + m_{2} = 1} & = 0 \,, \\
\mathcal{\omega}^{(o,+)}_{m_{1} + m_{2} = 1} & = - \frac{C_{1}R_{y}}{8\sqrt{2}Q_{5}} \Delta_{2,1,n_{1}+n_{2}}  \left( \frac{b_{1,m_{1},n_{1}}b_{1,m_{2},n_{2}}}{ 1+\epsilon_{1,2}} \right) \\
& \qquad \left\lbrace\sin v_{2,1,n_{1}+n_{2}}\left( \frac{a^{2}(n_{1}+n_{2})+2(1+n_{1}+n_{2})r^{2}}{a^{2}+r^{2}} \frac{dr}{r} +2 ((n_{1}+n_{2})\cos 2\theta - 2\sin^{2}\theta) \frac{d\theta}{\sin 2\theta} \right) \right. \\
& \qquad \qquad \left.+ \frac{1}{\Sigma} \cos v_{2,1,n_{1}+n_{2}} \left( (a^{2}(2+n_{1}+n_{2})\sin^{2}\theta +(n_{1}+n_{2})(a^{2}+r^{2}))d\varphi_{1}  \right.\right. \\
& \qquad\qquad\qquad\qquad\qquad\qquad\qquad\qquad \left. \left. +(a^{2}(n_{1}+n_{2})\cos^{2}\theta - (2+n_{1}+n_{2})r^{2})d\varphi_{2} \right)  \right\rbrace\,.
\end{align*}
Finally, when $m_{1} + m_{2} = 2$, the solution is given by:
\begin{align*}
    \mathcal{F}^{(o,+)}_{m_{1} + m_{2} = 2} &= 0\,,\\
    \mathcal{\omega}^{(o,+)}_{m_{1} + m_{2} = 2} &= - \frac{C_{1}R_{y}}{4\sqrt{2}Q_{5}} (2+n_{1}+n_{2})d_{2,2,n_{1}+n_{2}} \Delta_{2,0,2+n_{1}+n_{2}}\cot^{2}\theta  \\
    & \qquad \qquad \qquad \left[  \sin v_{2,2,n_{1}+n_{2}} \left( \frac{dr}{r} + \left( 1+\frac{a^{2}}{r^{2}}\right)\cot \theta \, d\theta \right)- \left( 1+\frac{a^{2}}{r^{2}}\right)  \cos v_{2,2,n_{1}+n_{2}} \, d\varphi_{1}  \right] \\
    & \qquad - \frac{R_{y}}{2\sqrt{2}}(d_{2,2,n_{1}+n_{2}}-b_{1,1,n_{1}}b_{1,1,n_{2}})\Delta_{2,2,n_{1}+n_{2}} \\
    & \qquad\qquad\qquad\qquad\left[ \frac{2}{a^{2}\sin 2\theta} \sin v_{2,2,n_{1}+n_{2}}\, d\theta - \frac{1}{\Sigma} \cos v_{2,2,n_{1}+n_{2}}\left( \left( 1+\frac{r^{2}}{a^{2}}\right)d\varphi_{1} - \frac{r^{2}}{a^{2}}d\varphi_{2} \right) \right]\,.
\end{align*}

\subsection{$S_{1,2}^{(s,+)}$ solutions}
\label{appSS:S12s+}

When solving the second layer with the ``addition mode" supercharged basic source:
\begin{align*}
\mathcal{S}_{1}^{(s,+)} & = \frac{R_{y}}{2}\left[\left(\frac{C_{1}}{ Q_{5}}+\frac{1}{\Sigma} \right)f_{2,m_{1}+m_{2},n_{1}+n_{2}}  +\frac{1}{ 2\Sigma}   (n_{2}-n_{1})(m_{2}-m_{1}) b_{1,m_{1},n_{1}}b_{1,m_{2},n_{2}} \right] \vartheta^{(s)}_{2,m_{1}+m_{2},n_{1}+n_{2}}  \,, \\
\mathcal{S}_{2}^{(s,+)} & = 0 \,,
\end{align*}
one need only consider the case when $m_{1}+m_{2} = 1$, due to (\ref{SuperchargedModeRestriction}).  

The solution to this source can not be found in the literature, but instead has to be solved for directly. Implementing a solution method directly paralleling that of \cite{Bena:2017xbt}, one can find the solution:
\begin{align*}
    \mathcal{F}^{(s,+)}_{m_{1}+m_{2}=1} &= 0 \,, \\
     \mathcal{\omega}^{(s,+)}_{m_{1}+m_{2}=1} &= - \frac{C_{1}R_{y}}{4\sqrt{2}Q_{5}}f_{2,m_{1}+m_{2},n_{1}+n_{2}} \Delta_{2,m_{1}+m_{2},n_{1}+n_{2}} \\ 
     & \qquad \left[ \sin v_{2,m_{1}+m_{2},n_{1}+n_{2}}\left(\frac{a^{2}}{a^{2}+r^{2}}\frac{dr}{r}+ \frac{2}{\sin 2\theta} d\theta \right) +\cos v_{2,m_{1}+m_{2},n_{1}+n_{2}}(d\varphi_{1}+d\varphi_{2}) \right]\,,
\end{align*}
provided the imposition of the coiffuring constraint:
\begin{equation}\label{CoiffApp1}
    f_{2,m_{1}+m_{2},n_{1}+n_{2}} = - \frac{1}{2}(m_{2}-m_{1})(n_{2}-n_{1})\left( \frac{b_{1,m_{1},n_{1}}b_{1,m_{2},n_{2}}}{1+\epsilon_{1,2}}\right)\,.
\end{equation}

\subsection{$S_{1,2}^{(-)}$ solutions}
\label{appSS:S12+}

Consider solving the second layer with the ``subtraction mode" basic source:
\begin{align*}
\mathcal{S}_{1}^{(-)} & = \f{R}{\sqrt{2} \Sigma} \Delta_{2,m_1+m_2,n_1+n_2} b_{1,m_1,n_1}b_{1,m_2,n_2}   \\
& \qquad \left\lbrace \left[(m_1-m_2+n_1-n_2)r\sin\theta\vphantom{\f{R}{2}}+ ((m_1-1)n_1 - (m_2-1)n_2)\f{\Sigma}{r\sin\theta}\right]\Omega^{(1)}\sin v_{0,m_1-m_2,n_1-n_2} \right.\\
& \qquad\qquad\left.+ \left[(m_1(1+n_1)+m_2(1+n_2))\Omega^{(2)} + ((m_1-1)n_1 + (m_2-1)n_2)\Omega^{(3)}\right]\cos v_{0,m_1-m_2,n_1-n_2}\right\rbrace \,, \\
\mathcal{S}_{2}^{(-)} & =  \frac{1}{R_{y}}  \Delta_{2,2m_{1},2n_{1}}b_{1,m_{1},n_{1}}b_{1,m_{2},n_{2}} \\
& \qquad \left\lbrace \frac{1}{\Sigma}(m_{2}-m_{1}+n_{2}-n_{1})^{2} + 2\left[\frac{(1-m_{1})(1-m_{2})n_{1}n_{2}}{r^{2}\sin^{2}\theta} + \frac{m_{1}m_{2}(1+n_{1})(1+n_{2})}{(a^{2}+r^{2})\cos^{2}\theta} \right]    \right\rbrace  z^{(o)}_{0,m_{2}-m_{1},n_{2}-n_{1}} \,.
\end{align*}
Since $C_{1}$ does not appear in this source, the solution can be obtained directly from the asymptotically AdS $(1,m,n)$ solution constructed in \cite{Mayerson:2020tcl}: 
\begin{align*}
    \mathcal{F}^{(-)} &= \frac{1}{a^{2}}b_{1,m_{1},n_{1}}b_{1,m_{2},n_{2}} \left( \delta_{m_{1}+m_{2},0}\Delta_{0,0,n_{1}+n_{2}}+\delta_{m_{1}+m_{2},2}\delta_{0,0,2+n_{1}+n_{2}} \right) \cos v_{0,m_{2}-m_{1},n_{2}-n_{1}}\,, \\
     \mathcal{\omega}^{(-)} &= \frac{R_{y}}{\sqrt{2}a^{2}}b_{1,m_{1},n_{1}}b_{1,m_{2},n_{2}}\Delta_{2,m_{1}+m_{2},n_{1}+n_{2}} \left[ (m_{2}-m_{1})\sin v_{0,m_{2}-m_{1},n_{2}-n_{1}} \frac{d\theta}{\sin 2\theta} \right.\\
     & \qquad  \left.- \frac{1}{2\Sigma}\left( (a^{2}+r^{2})(2\delta_{m_{1}+m_{2} , 0}+\delta_{m_{1}+m_{2},1})d\varphi_{1} -(\delta_{m_{1}+m_{2},1}+2\delta_{m_{1}+m_{2},2})r^{2}d\varphi_{2} \right)\cos v_{0,m_{2}-m_{1},n_{2}-n_{1}} \right]\,.
\end{align*}

\subsection{Aggregating solution}
\label{appSS:FullSol}
Given the form of the second layer source decomposition (\ref{SourceDecomp}):
\begin{align} 
\mathcal{S}_{1,2} &= \sum_{m_{1},m_{2}}\sum_{n_{1},n_{2}} \left(\mathcal{S}_{1,2}^{(-)}+\mathcal{S}_{1,2}^{(o,+)} + \mathcal{S}_{1,2}^{(s,+)} \right) \,.
\end{align}
Utilizing the linearity of the second BPS layer, the results of the preceding subsections can be summed, to give the full second layer solution. This results in a solution for $\mathcal{F}$ of the form:
\begin{align}
\mathcal{F} &= \sum_{m_{1},m_{2}}\sum_{n_{1},n_{2}} \mathcal{F}^{(-)}\,,
\end{align}
and a solution for $\omega$ of the form: 
\begin{align} 
\omega &= \sum_{m_{1},m_{2}}\sum_{n_{1},n_{2}} \left(\omega^{(-)}+\omega^{(o,+)} + \omega^{(s,+)} \right) \,.
\end{align}
Where $(\mathcal{F}^{(-)} ,\omega^{(-)}, \omega^{(o,+)} , \omega^{(s,+)} )$, for the various $(m_{1},m_{2})$ allowed in the $(1,m,n)$ superposition, are presented in Sections \ref{appSS:S120+} through \ref{appSS:S12+}.

Further, in constructing these solutions, the coiffuring constraints (\ref{CoifdApp1}) and (\ref{CoiffApp1}):
\begin{align}
    d_{2,m_{1}+m_{2},n_{1}+n_{2}} &= \frac{b_{1,m_{1},n_{1}}b_{1,m_{2},n_{2}}}{ 1+\epsilon_{1,2}} \,,\\
    f_{2,m_{1}+m_{2},n_{1}+n_{2}} &= - \frac{1}{2}(m_{2}-m_{1})(n_{2}-n_{1})\left( \frac{b_{1,m_{1},n_{1}}b_{1,m_{2},n_{2}}}{1+\epsilon_{1,2}}\right)\,,
\end{align}
must be implemented. The ultimate validity of these coiffuring relations comes from showing that the final geometry is regular, which is demonstrated/argued in the body of this paper.

\begin{adjustwidth}{-1mm}{-1mm} 
\bibliographystyle{JHEP}      
\bibliography{AsyFlat}       

\end{adjustwidth}


\end{document}